\documentclass[12pt,preprint]{aastex}

\slugcomment{}

\shorttitle{Broad Line Region in NGC 4203}
\shortauthors{Devereux et al.}

\begin{document}


\title{Time Variable Broad Line Emission in NGC 4203: Evidence for Stellar Contrails}


\author{Nick Devereux}
\affil{Department of Physics, Embry-Riddle Aeronautical University,
    Prescott, AZ 86301}
\email{devereux@erau.edu}



\begin{abstract}

Dual epoch spectroscopy of the lenticular galaxy, NGC 4203, obtained with the ${\it Hubble~Space~Telescope~(HST)}$ has revealed that the double-peaked
component of the broad H${\alpha}$ emission line is time variable, increasing by a factor of 2.2 in brightness between 1999 and 2010. Modeling the gas distribution responsible for the double-peaked profiles indicates that a ring is a more appropriate description than a disk and most likely represents the contrail of a red supergiant star that is being tidally disrupted at a distance of ${\sim}$ 1500 AU from the central black hole.
There is also a bright core of broad H${\alpha}$ line emission that is not time variable and identified with a large scale inflow from an outer radius  ${\sim}$ 1 pc.
If the gas number density is ${\geq}$ 10$^6$ cm$^{-3}$, as suggested by the absence of similarly broad [O I] and [O III] emission lines, then the steady state inflow rate is ${\sim}$ 2 ${\times}$ 10$^{-2}$ M${_{\sun}}$/yr which exceeds the inflow requirement to explain the X-ray luminosity in terms of radiatively inefficient accretion by a factor of ${\sim}$ 6.
The central AGN is unable to sustain ionization of the broad line region, the discrepancy is particularly acute in 2010 when the broad H${\alpha}$ emission line
is dominated by the contrail of the in-falling supergiant star. However, ram pressure shock ionization
produced by the interaction of the in-falling supergiant with the ambient interstellar medium may help alleviate the ionizing deficit
by generating a mechanical source of ionization supplementing the photoionization provided by the AGN.

 \end{abstract}





\section{Introduction}

NGC 4203 is a lenticular galaxy located at a distance of 15.14 Mpc \citep{Ton01}. The galaxy is of interest because it harbors a low ionization nuclear emission-line region \citep[LINER;][]{ Hec80}
of type 1.9 indicating that it exhibits a broad H${\alpha}$ emission line \citep{Ho97a, Ho97b}.
High angular resolution observations with the {\it Space Telescope Imaging Spectrograph (STIS)}
aboard the {\it Hubble Space Telescope} ({\it HST}) revealed that the broad H${\alpha}$ emission line is actually double-peaked \citep{Shi00} with a FWZI of $\sim$ 12,000 km/s. 
NGC 4203 was observed with {\it STIS} again 11 years later revealing a significantly brighter double-peaked broad 
H${\alpha}$ emission line. Thus, NGC 4203 joins a short but growing list of LINERs with time variable broad Balmer emission lines 
of which the best studied nearby example is NGC 1097 \citep{Sto93, Sto95, Sto97} but also includes M81 \citep{Bow96, Dev03} and NGC 3065 \citep{Era01}. 
NGC 4203 also hosts a time variable, compact core, flat-spectrum radio source \citep{Nag02, Ulv01}.  \cite{Mao05} showed that the nucleus of NGC 4203 is also 
variable in the ultra-violet (UV). However, variability has yet to be established in X-rays \citep{Pia10, You11}. \cite{Mao07} compiled the radio-to-X-ray spectral energy distribution of the active galactic nucleus (AGN) showing that it resembles more luminous Seyferts. Several attempts have been made to estimate the black hole (BH) mass using spatially resolved gas kinematics \citep{Shi00, Sar01, Sar02, Bei09} with mixed results. Subsequently, \cite{Lew06} adopted the stellar velocity dispersion measurement of 
\cite{Bar02} to estimate a BH mass of (6 ${\pm~1) \times}$ 10${^7}$ M${_\odot}$, which, when compared to the bolometric luminosity, indicates that NGC 4203 radiates well below, $\sim$10$^{-5}$, of the Eddington luminosity limit \citep{Lew06,Mao07}.   

Accretion disks are expected to produce broad emission lines with characteristics that, in principle, betray their geometries \citep{Lew10, Gez07}. 
For example, disks of low inclination are predicted to 
produce an asymmetric profile with the blue peak brighter than the red \citep{Che89}. As the disk becomes more edge-on, the blue and red peaks achieve a similar brightness and may even merge into a single peak if the disk has a large ratio of outer radius to inner radius as may be the case in NGC 3065 \citep{Era01}. However, disk models invariably are unable to simultaneously explain both the broad shoulders
and the bright core of broad line emission often seen centered at the systemic velocity of the galaxy \citep{Dow10, Gez07, Era94}, especially in nearby galaxies observed with high angular resolution. For example, in NGC 4579, NGC 4450 and NGC 4203,
\cite{Bar01}, \cite{Ho00} and \cite{Shi00}, respectively, attribute the central core emission to a {\it ``normal"}  broad component. 
Similarly, \cite{Dow10} and \cite{Gez07} ascribe the broad H${\alpha}$ emission lines seen in distant quasars and broad line radio galaxies in terms of an accretion disk and a 
{\it ``standard"} or {\it ``classical"} broad line region (BLR). However, labeling the single peak of core emission as the {\it ``normal"}, {\it ``standard"} or {\it ``classical"}  BLR does not necessarily entail an explanation of what it is exactly. There are two possibilities; it is either an inflow or an outflow.
Radiatively driven outflows are not possible, however, in LINERs, because they radiate well below the Eddington luminosity limit. Thus, 
inflow is the only remaining explanation for the central core emission seen in LINERs with double-peaked profiles. 

This paper is the third in a
series that attempts to model the broad emission line profiles for low luminosity AGNs (LLAGNs) with known central masses and derive a size for the BLR.
Inflows can produce the single peaked broad emission line profiles seen in M81 and NGC 3998 \citep{Dev07,Dev11} and inflows are expected to 
be associated with accretion disks, because the former, presumably, fuels that latter.
An objective of this paper, therefore, is to demonstrate the utility of an {\it inflow plus disk model} to explain 
double peaked broad emission line profiles with central core emission and, by way of illustration, apply this combined model to interpret the time-variable broad 
Balmer emission lines seen in NGC 4203. The wider context for this investigation is to better understand the 
broad line phenomenon in LLAGNs by characterizing their physical properties using archival {\it HST-STIS} spectroscopy. NGC 4203 is an excellent candidate to study as it is nearby, bright and radiates well below the Eddington luminosity limit and is therefore unable to sustain a radiatively driven outflow thereby simplifying the interpretation of the broad emission lines by eliminating one of the possible explanations for their origin. 

Since the BLR is unresolved, the spatial distribution of emission cannot be directly measured.
However, the three dimensional gravitational field strength is known, by virtue of the BH mass, thus the relationship
between velocity and radius may be established, given a kinematic model for the broad emission line gas.
In this way, one can, in principle, exploit the exquisite velocity resolution of ${STIS}$ to model the broad emission line profiles and determine a size for the BLR that heretofore has previously only been possible using reverberation mapping techniques \citep{Pet93,Pet01}.
Profile fitting complements reverberation mapping as the latter yields BLR sizes for a different sample of more
luminous AGNs. 

The layout of the paper is as follows. In Section 3, the BELs seen in NGC 4203 are evaluated in the context of
an inflow plus disk model. Some physical properties of the BLR are presented in Section 4. Conclusions follow
in Section 5. We begin, however, with Section 2 and a review of the emission lines observed in the nucleus of NGC 4203.

\section{NGC 4203 Emission Lines }


NGC 4203 has been observed with {\it STIS}  twice using the G750M grating and once using the G430L grating. Table 1 summarizes the observations and the archival spectra are presented in Fig 1. 
A {\it STIS} spectrum showing the broad H${\alpha}$ emission line has been presented previously
by \cite{Shi00, Shi07} based on observations obtained under PID 7361. 
 \cite{Shi00}, \cite{Bas03}, \cite{GD04} and \cite{Sar05} have all presented the G430L spectrum obtained under PID 7361. However, the G750M spectrum obtained under PID 11571 is shown for the first time in Fig. 1. 

Multiple exposures obtained using the same grating and at the same position have been combined using the {\bf STSDAS} task {\bf ocrreject}.
Dithered exposures were shifted using the {\bf STSDAS} task {\bf sshift} prior to combining. Subsequently, the {\bf STSDAS} task {\bf x1d} was used to perform a 7 pixel wide extraction along the slit direction and centered on the nucleus. Each extraction samples ${\ge}$ 80\% of the encircled energy for an unresolved point source \citep{Pro10}.

The G750M spectra reveal a broad H${\alpha}$ line with a FWHM ${\sim}$ 3000 km/s that brightened by more than a factor of two in the almost 11 year interval since the first spectrum was obtained. Two narrow but resolved [S II] lines appear on the red side of the broad H${\alpha}$ line at ${\sim}$ 6750{\AA} and two narrow but resolved [O I] lines appear on the blue side at ${\sim}$ 6350{\AA}. These narrow lines 
are virtually identical in the two datasets. Thus, the brightening appears to be limited to just the broad H${\alpha}$ + [N II] emission line complex which is remarkable because the more recent observations were obtained with a smaller 0.1{\arcsec} slit\footnotemark \footnotetext{The continuum did change slightly, down by 15\% compared to 1999 in the sense that the 2010 continuum is slightly redder but it is not clear whether this difference is due to the different slit sizes used for the observations or whether it reflects an intrinsic change in the AGN.}.
The G430L spectrum reveals a wide swath of emission lines. Collectively, the G750M and G430L spectra resolve the H${\alpha}$, H${\beta}$ and H${\gamma}$ lines.
A more detailed description of the spectra follows, beginning with the Hydrogen (H) Balmer lines.

\subsubsection{Broad H${\alpha}$ Emission }

The H${\alpha}$ emission line profile exhibits broad shoulders that are accentuated in the 2010 spectrum compared to the one obtained in 1999.  Both of the [N II] vacuum wavelength 6549.85 and 6585.28 {\AA} emission lines and the narrow component of the vacuum wavelength 6564.61 H${\alpha}$ emission line can be seen in a 7 pixel extraction centered on the nucleus in both datasets which greatly facilitates their modeling. The ${\bf STSDAS}$ contributed task ${\bf specfit}$ was used to model and subtract the forbidden lines and the results are reported
in Table 2 along with the fluxes for the two [S II] vacuum wavelength ${\lambda}$6718.29 and ${\lambda}$6732.67 {\AA} emission lines
and the two [O I] vacuum wavelength ${\lambda}$6302.04 and ${\lambda}$6365.53 {\AA} emission lines. 
The flux for the model [N II] lines and the model ${\it narrow}$ H${\alpha}$ line has been chosen so that the difference spectrum does not show an `over-subtraction' of the broad H${\alpha}$ profile. The procedure is valid because the 
observed broad H${\alpha}$ emission line profile is otherwise smoothly varying. 
In support of the procedure, the model parameters deduced for the [N II] lines are consistent with the expectation based on atomic physics that sets the wavelength of the fainter [N II] line relative to the brighter [N II] line,
constrains the flux of the fainter [N II] line to be 1/3 that of the brighter [N II] line, and requires that the
[N II] lines share the same width. Two components were required to model the brightest [N II] line in order to obtain a satisfactory subtraction
as noted in Table 2. A single gaussian was used to model the other lines.
The brightest [O I] line is wider than the brightest [N II] line at the ${\sim}$ 90\% confidence level. That these forbidden
lines have similar widths and further that the narrow H${\alpha}$ line is fainter than the brightest
[N II] line are characteristics shared by other AGNs observed with {\it STIS} for which these lines can be clearly seen \citep{NS03}. 
A model narrow line spectrum is illustrated in Fig. 2 along with the difference spectrum which reveals a broad H${\alpha}$ line profile that is very similar to the 
H${\beta}$ line described in the next section.
Figure 2 illustrates that the consequence of subtracting the forbidden emission line model is to reveal a broad
H${\alpha}$ emission line symmetric about the ${\lambda}$6588{\AA} wavelength expected for a systemic velocity of 1077 km/s deduced from the peak of the brightest [N II] emission line. 
Thus, there is no apparent redshift between the broad H${\alpha}$ emission line and the systemic redshift of the host galaxy. The broad H${\alpha}$ emission
line flux deduced from the 1999 G750M spectra is ${\sim}$ 40\% larger than estimated previously by \cite{Shi00}. The difference can be attributed to the fact that \cite{Shi00} subtracted not only the forbidden emission lines described above, but also another component that they call the {\it ``normal"} broad line emission centered at the systemic velocity of the galaxy. Comparing the entire H${\alpha}$ emission line flux measured in the 2010 spectrum with that measured for the 1999 spectrum indicates that the broad line increased in brightness by a factor of 2.2.

\subsubsection{Broad H${\beta}$ and H${\gamma}$ Emission }

The H${\beta}$ and H${\gamma}$  emission lines are compromised only slightly by [O III] emission lines
which were modeled and subtracted using the
${\bf STSDAS}$ contributed task ${\bf specfit}$ as illustrated in Fig. 3 and Fig. 4. The [O III] vacuum wavelength ${\lambda}$4364.44 {\AA}
deserved special attention as it is very faint and marginally resolved.  A model was employed
for the [O III] ${\lambda}$4364.44 line in which the wavelength and line width was fixed 
and the flux adjusted so as to not over-subtract the H${\gamma}$ line in the difference spectrum (Fig. 4). This yielded the upper limit for the [O III] ${\lambda}$ 4364.44 line reported in Table 3.  Subtracting the [O III] lines reveals broad H${\beta}$ and H${\gamma}$ emission lines.
Unfortunately, the analysis can not reliably be extended to include other Balmer lines as they are even fainter and even more confused with emission from other ions.
Consequently, emission line fluxes are reported in Table 3 for all the lines that can be reliably resolved and measured in the G430L spectrum
including H${\beta}$, H${\gamma}$, the blend of the vacuum wavelength ${\lambda\lambda}$3727.09, 3729.88 {\AA} [O II] lines,  the two [O III] vacuum wavelength ${\lambda}$4960.30 {\AA} and ${\lambda}$5008.24 {\AA} emission lines, and the 
upper limit for the vacuum wavelength ${\lambda}$4364.44 {\AA}  [O III] emission line. According to the diagnostic diagram of \citet[][their Fig. 5]{Kew06} 
the [O III] ${\lambda}$5008.24 /[O II]  ${\lambda\lambda}$3727.09, 3729.88 and [O I] ${\lambda}$6324.99/H${\alpha}$ emission line ratios measured with {\it STIS} in 1999 qualify NGC 4203 as a Seyfert and the difference between the permitted and forbidden line widths further qualifies NGC 4203 as a Seyfert 1.

\subsubsection{Similar Balmer Line Profiles}

Figure 5 illustrates the striking similarity between the H${\alpha}$ and H${\beta}$ lines in NGC 4203 when they are normalized to their respective peak intensities and the wavelength scales converted 
to velocity using the non-relativistic Doppler equation. The similarity between the emission line profiles is particularly impressive considering the variety of models employed to subtract the superimposed forbidden lines. On the other hand, the H${\gamma}$
line profile is quite different in the sense that the line core appears to be broader and the broad shoulders are absent, 
particularly noticeable on the blue side. It is not clear whether the difference between the H${\gamma}$ profile and the other
Balmer lines is a consequence of lower signal-to-noise data for this weak line or whether it represents a real radiative transfer effect. 
Nevertheless, the broad shoulders on the H${\alpha}$ and H${\beta}$ lines suggest an accretion disk,
as noted previously by \cite{Shi00}, whereas the line core is similar to what would be expected 
for a spherically symmetric shell of gas in radial motion as noted 
previously by \citet{Dev07} and \cite{Dev11}. Each of these models will be described further in Section 3.

\subsubsection{Balmer Decrements}

Despite the similarity between the broad H${\alpha}$ and H${\beta}$ line profiles, the Balmer decrement, 
H${\alpha}$/H${\beta}$ = 5.0 ${\pm}$ 0.1 is significantly different from the Case B value, 2.75, 
in the sense that the observed value is systematically 182\% higher. Interpreting this ratio in terms of dust extinction leads to a color excess 
$E(B - V)$ ${\sim}$ 0.6 mag which is inconsistent with the reddening estimated to the AGN X-ray emission by \cite{Pia10}. Such deviations from recombination theory have been noted for other LINERs \citep[e.g.][]{Dev11,Sto97,Bow96,Fil84} and have been attributed to collisional excitation in gas of high density or time variable dust extinction.




\section{The Inflow Plus Disk Model}

As alluded to in the Introduction and in Section 2.0.3, the broad H${\alpha}$ (and H${\beta}$) line profiles can be modeled using the combination
of an accretion disk {\it and} an inflow\footnotemark  \footnotetext{A radiatively driven outflow of gas can be ruled out for NGC 4203 because the diminutive luminosity generated by the AGN is simply unable to provide sufficient radiation pressure to overcome the gravitational force of the BH.} since neither model alone is able to explain the entire broad emission line. The disk model, described previously by \cite{Che89}, describes the emission expected from a thin axisymmetric, relativistic accretion disk. The model includes self-consistently all relevant relativistic effects, such as Doppler boosting and transverse and gravitational redshifts and involves five free parameters, a
dimensionless inner radius, ${\xi}_{i}$, and outer radius,
${\xi}_{o}$, an inclination angle measured from the disk normal to the line
of sight, $i$, an emissivity law of the form
$\varepsilon\propto r^{-q}$, and
a velocity dispersion for the gas, ${\sigma}$ in km/s. The inflow model has been described most recently by \cite{Dev11}. It describes
a steady state spherically symmetric infall and invokes just two free parameters, an inner and outer radius; $r_{inner}$ and $r_{outer}$, respectively.
Collectively, the disk plus inflow model encompass 8 free parameters if one includes the scaling that determines the relative contribution of the disk and inflow to the total line emission (see Tables 4 \& 5).

Illustrative models presented in Fig. 6 show that the broad H${\alpha}$ line profile observed in 1999 is dominated by the inflow with only a small contribution from the disk, except at the extremities of the line. However, the situation is quite different for the broad H${\alpha}$ line profile observed in 2010 which, according to the model,
appears to be dominated by the accretion disk, except in the line core. The inflow model parameters are chosen to be the same for the 1999 and 2010 data, motivated by the fact that the narrow lines, including the narrow component of the H${\alpha}$ line, did not change between 1999 and 2010. Thus, only the disk changed. Experimentation revealed that the 2010 H${\alpha}$ line profile is best modeled with a ratio of outer to inner disk radius, ${\xi}_{o}$/${\xi}_{i}$ = 3.75, a velocity dispersion, ${\sigma}$ = 350 km/s and an
emissivity index, q=3.
A more thorough exploration of the parameter space was performed by computing the value of ${\chi}_{red}^2$ for a grid of 156 models per epoch spanning 26 inclinations\footnotemark  \footnotetext{Inclinations smaller than 40${^\circ}$ yielded no ${\chi}^2_{min}$ = 1 solutions for the 2010 observations.} between 40${^\circ}$ ${\le}$ ${i}$ ${\le}$ 90${^\circ}$, and 6 values for the inner radius spanning 500 ${\le}$ ${\xi}_{i}$  ${\le}$ 3000, expressed in units of the gravitational radius, ${r_g}$ = GM${_{\bullet}}/c^2$. The remaining parameters are fixed as listed in Table 4.

\begin{equation}
{\chi}_{red}^2 = {\sum_j} (O_j - M_j)^2/(\nu \delta^2)
\end{equation}

where ${O_j}$ represents the observed normalized line profile intensities,  ${M_j}$, the model normalized line profile intensities, ${\nu}$ is the number of degrees of freedom
and ${\delta}$ the uncertainty in the observed normalized line profile intensities, taken to be 4\% for both epochs. The summation was performed over the velocity span of the broad H${\alpha}$ line as specified in Table 5.  The minimum ${\chi}_{red}^2$ computed for the 2010
data was unity indicating that the axis-symmetric disk plus inflow model provided an excellent representation of those data.  A contour plot
of the ${\chi}_{red}^2$ surface, presented in Figure 7, shows that there are multiple combinations of disk inclination and inner radius that can minimize ${\chi}_{red}^2$ for each epoch. Thus, the model representations of the data presented in Fig. 6 are by no means unique. However, the fact that the red and blue ${\chi}^2_{min}$ contours do not overlap suggests that there is no single disk model that can simultaneously explain both the 1999 and 2010 observations. 
There could even be two disks. But, the most conservative interpretation is that there is just one disk for which the inclination did not change between 1999 and 2010. In this
case the size of the disk must have changed. The arrow in Fig. 7 represents the ${\it least}$ size evolution solution which implies
that the disk is highly inclined, ${i}$ = 82${^\circ}$ ${\pm}$ 1${^\circ}$, which presumably aided in its detection, and the inner radius of the disk decreased from 2770${r_g}$ ${\pm}$ 182${r_g}$ in 1999 to 2200${r_g}$ ${\pm}$ 216${r_g}$ in 2010, the implications of which are discussed further in Section 4.7.

\section{Discussion}

\subsection{BLR Size}

That the inflow plus disk model is able to reproduce the rather complex shapes observed for the broad H${\alpha}$ emission line profiles in NGC 4203 allows
the sizes of the inflow and disk components to be estimated. The mass 
distribution determines that the size of the inflow is large;  ${\sim}$ 2 pc ( 27 milli-arcsec) in diameter\footnotemark \footnotetext{With an angular diameter of 0.02${\arcsec}$ the BLR of NGC 4203 is spatially unresolved with {\it STIS}.} and it terminates at the much smaller ${\sim}$ 14 ${\times}$ 10${^{-3}}$ pc (200 ${\mu}$arcsec) diameter disk.

The physical size inferred for the inflow and the disk components causes NGC 4203 to not conform to the
correlation between BLR size and UV luminosity established for quasars and high
luminosity AGNs using reverberation mapping \citep{Pet01,Pet93, Kas05}. 
An extrapolation of the BLR size - luminosity relationship\footnotemark \footnotetext{FITEXY at 1450{\AA}.} of \cite{Kas05} down to the low UV luminosity estimated for the AGN in NGC 4203, predicts a radius for the BLR that is ${\sim}$ 0.5 l.d, which is ${\sim}$ 20 times smaller than inferred for the inner radius of the disk using line profile fitting (Section 3). Conversely, the large outer radius determined for the inflow in NGC 4203 using profile fitting makes it physically larger than any BLR measured using reverberation mapping, 4 times larger than the previous record holder; the quasar 3C 273 \citep{Kas05}. 
However, as
noted by \cite{Kas05}, the BLR size -- luminosity correlation appears to break down
for low luminosity AGNs, which, with L(2500 ${\textrm \AA}$) = 4.4 ${\times}$ 10${^{40}}$ erg/s \citep{Mao07}, would include NGC 4203. 
Of course, the BLR size -- luminosity relationship of \cite{Kas05} is defined by quasars and AGNs that are orders of magnitude more luminous than NGC 4203. Thus, the very large discrepancy arising from the comparison strongly suggests that the BLR in NGC 4203 is not simply that of a scaled down quasar.

\subsection{Virial Black Hole Masses}

Estimating the mass of the BH in NGC 4203 using the so called `virial method'  leads to a value that is 
substantially lower than the mass estimated from the stellar velocity dispersion by \cite{Lew06}. For example, the formalism of \cite{Gre05}, which uses the FWHM {\it and} luminosity of the broad H${\alpha}$ emission line, underestimates the mass of the BH in NGC 4203 by a factor of 348 using the 
broad H${\alpha}$ emission line observed in 1999 and a factor of 51 using the one observed in 2010.
This dichotomy is regarded as further evidence that the BLR  in NGC 4203 is very different from the BLR in more luminous AGNs.

\subsection{Broad Line Region Ionization}

Following \cite{Mao07}, the ionizing continuum generated by the AGN in NGC 4203 may be represented as a power law
with an optical to X-ray spectral index ${\alpha}$ = 1.11 and a normalization provided by the UV continuum described in Section 4.1. 
Integrating the spectral energy distribution from 13.6 eV to 100 keV using the method described previously
in \cite{Dev11} predicts 1.7 x ${10^{51}}$ ionizing ph/s after correcting for 0.05 mag of Galactic extinction only. For comparison, the broad  H${\alpha}$ emission line flux measured in 1999 (Table 2) which is dominated by the inflow (Fig. 6) corresponds to an H${\alpha}$ luminosity,  $L(H{{\alpha}}$) = 1.14 ${\times}$10$^6$ L${_{\sun}}$, which, assuming 45\% of the ionizing photons are converted into H${\alpha}$ photons (Case B recombination at a temperature of 10$^4$ K) requires (3.2 $\pm$ 0.04) x ${10^{51}}$ ionizing ph/s using,

\begin{equation}
N_{ion} = L(H{_{\alpha}}){\alpha_{B}}/{\alpha^{eff}_{H\alpha}} h \nu_{H\alpha} 
\end{equation}

where ${\alpha^{eff}_{H\alpha}}$ = 1.16 x 10$^{-13}$ cm${^{3}}$ s${^{-1}}$ is the effective recombination coefficient  and 
${\alpha_{B}}$ = 2.59 x 10$^{-13}$ cm${^{3}}$ s${^{-1}}$ is the total Case B recombination coefficient.

Thus, the number of ionizing photons required to excite the broad H${\alpha}$ emission line measured in 1999 exceeds the number of ionizing photons available from the AGN by a factor ${\sim}$ 2, independent of the gas density and the filling factor. The discrepancy becomes larger for the brighter H${\alpha}$ emission line measured in 2010. Then the number of ionizing photons required to excite the H${\alpha}$ emission line exceeds that available from the AGN by a factor of ${\sim}$ 4. Technically, these are upper limits because the AGN is variable in the UV 
and could have brightened prior to and since 2003 when it was last measured \citep{Mao05}\footnotemark \footnotetext {The ionizing deficit increases by another factor of 5 if the extinction to the H${\alpha}$ emission line is estimated from the Balmer decrement mentioned previously in Section 2.0.4. On the other hand, the
anomalous Balmer decrement may be the consequence of high gas density as noted in Section 4.7.}. Nevertheless, tentatively, NGC 4203 joins M81 and NGC 4579 as examples of LINERs with a BLR {\it ionizing deficit} \citep{Dev07,Mao98,Ho96}.

\subsection{Constraints on the Inflow Gas Density}

The size of the aperture employed to extract the spectra illustrated in Fig. 1 encompasses emission from gas up to a radial distance of 13 pc from the central AGN
where the velocity dispersion is expected to be ${\sim}$ 200 km/s and comparable to that measured for the [S II] lines.
Thus, the two [S II]  lines provide an opportunity to measure the gas density well beyond the outer boundary of the inflow. 
The nominal value for the observed [S II] ${\lambda}$6742/${\lambda}$6757 
intensity ratio = 0.68 ${\pm}$ 0.1, corresponding to n = 2 ${\times}$ 10$^3$ cm$^{-3}$, with uncertainties permitting densities in the range 1 ${\leq}$ n (10$^3$ cm$^{-3}$)  ${\leq}$  6. The fact that the FWHM of the broad H${\alpha}$ and H${\beta}$ emission lines\footnotemark \footnotetext{Observed in 1999 and hence dominated by the inflow.} is larger than that of the [O I] and [O III] emission lines (Table 2 \& 3) means that the gas density ${\it inside}$ the inflow must be greater than the critical density of the levels from which the  [O I] and [O III] lines originate which is ${\sim}$10$^6$ cm$^{-3}$. 
This limit is higher than the one obtained from the [S II] line ratio implying that the gas density increases towards the central BH.

\subsection{The Mass of Ionized Gas Required to Produce the Broad H${\alpha}$ Emission Line}

The mass of emitting gas may be deduced from the broad H${\alpha}$ emission line 
luminosity assuming standard (Case B) recombination theory;

\begin{equation}
 M_{emitting} =  L (H\alpha) m_H /  \rm{n}  {\alpha^{eff}_{H\alpha}} h \nu_{H\alpha} 
\end{equation}

Using an effective recombination 
coefficient ${\alpha^{eff}_{H\alpha}}$ =
1.16 x 10$^{-13}$ cm${^{3}}$ s${^{-1}}$, assuming a \textit{constant} average density n ${\ge}$  10$^6$ cm$^{-3}$,
and a luminosity $L (H{{\alpha}}$) = 1.14 x 10${^6}$ L${_{\sun}}$ based on the broad line flux measured in 1999 (Table 2), leads to an upper limit on the mass of inflowing ionized gas\footnotemark \footnotetext{The brighter line in the 2010 spectrum approximately doubled the mass of ionized gas compared to that seen in 1999 if the number density is the same. Although, evidence presented in Section 4.7 suggests
that the number density for the gas producing the double-peaked broad emission line in 2010 is much higher.}, $M_{emitting}$ ${\le}$ 10 M${_{\sun}}$. 
If only a fraction of the gas is ionized then the upper limit on the  {\it total} (ionized + neutral) gas mass could, of course, be higher.

\subsection{An Assessment of the Inflow Scenario for NGC 4203}

It is straight forward to calculate the volume filling factor, ${\epsilon}$, for the ionized gas producing the H${\alpha}$ emission once the dimensions of the emitting region have been established. For a uniform density medium occupying a spherical volume of radius $r$, one finds 

\begin{equation}
 \epsilon = 3 L (H_\alpha)/ 4  \pi  \rm{n}_H^2  {\alpha^{eff}_{H\alpha}} h \nu_{H\alpha} r^3
\end{equation}

Again, using an effective recombination 
coefficient ${\alpha^{eff}_{H\alpha}}$ = 1.16 x 10$^{-13}$ cm${^{3}}$ s${^{-1}}$, assuming a \textit{constant} average number density n ${\ge}$ 
10$^6$ cm$^{-3}$, and a luminosity $L (H{_{\alpha}}$) = 1.14 x 10${^6}$ L${_{\sun}}$ based on the 1999 measurement of the broad line flux reported in Table 2, leads to an upper limit on the  filling factor, ${ \epsilon}$ ${\le}$ 1 x 10$^{-4}$, for NGC 4203 if 
the size of the BELR, $r$ = 1 pc. The very low filling factor suggests that the inflow is not continuous but
composed of many ionized, density bounded, gas filaments. Such filaments would have approximately the same gas density and hence the same emissivity regardless of their location with respect to the central AGN, they would be optically thin and hence emit isotropically. Ionized gas filaments embrace all the essential elements of the inflow model and are commonly seen in the nuclei of active galaxies \citep{Sto09} including the central parsec of the Galactic Center \citep{Lac91}. Consequently, the inflow modeled here is not a Bondi flow \citep{Bon52} because the inflowing `particles' are discrete and do not constitute a continuous fluid.

Having established the dimensions of the emitting region and a lower limit on the gas density one
can estimate the mass inflow rate, $\dot{m}$, for the ionized gas observed in 1999, using the equation of continuity;

\begin{equation}
 \dot{m}  =  \epsilon 4 \pi r^2 \rm{v}  \rm{n} m_H
\end{equation}

The free-fall velocity, v, at a radius of 1 pc is determined by the mass distribution to be 730 km/s. Setting ${ \epsilon}$ ${\le}$ 1 x 10$^{-4}$ for a self-consistent gas density in the flow
of n ${\ge}$10$^6$ cm$^{-3}$ (see section 4.4), one obtains a mass inflow rate,  $\dot{m}$ ${\sim}$ 2.4 ${\times}$ 10$^{-2}$ M${_{\sun}}$/yr. However, if only a fraction of the inflowing gas is ionized, then the {\it total} mass inflow rate could, of course, be higher.

The 1--10 keV X-ray luminosity adopted for the AGN in NGC 4203, $L_{1-10~keV}$, is $7.3\times 10^{40}~{\rm erg~s^{-1}}$
\citep{Pia10}. Assuming this is powered by radiatively inefficient accretion
leads to the following formula  \citep{Mer03}

\begin{equation}
 L_{1-10~ keV}= 7  \times 10^{38} M_{\bullet}^{0.97}  \dot{m}^{2.3}
\end{equation}

where $L$$\rm_{1-10~keV}$ is in ${erg~s^{-1}}$ and $M$$_{\bullet}$ is in solar masses. Under these
circumstances the accretion rate required to power the observed X-ray emission, $\dot{m} \sim 3.9 \times
10^{-3} ~{\rm M_\odot~yr^{-1}}$.  Thus, the steady state spherically symmetric inflow 
inferred from the broad H${\alpha}$ line emission seen in 1999 exceeds the requirement to explain the X-ray luminosity 
in terms of radiatively inefficient accretion by a factor of ${\sim}$ 6.

The most obvious source for the inflowing material is stellar-mass loss \cite[e.g.][]{Ho09} with the caveat that it is only the gas for which the vector cross product

\begin{equation}
r {\times} p = 0
\end{equation}

where r is the radius vector and p is the linear momentum of the gas, will be accreted by the BH. This is expected to limit the accretion rate to 
the values ${\leq}$ 0.1 ${\rm M_\odot~yr^{-1}}$ measured for the inflows in M81 \citep{Dev07}, NGC 3998 \citep{Dev11} and NGC 4203 studied here.

\subsection{An Assessment of the Accretion Disk Scenario for NGC 4203}

There are four indications that the accretion disk modeled in NGC 4203 is not, in fact, an accretion disk. First, the FWHM and the total broad line flux both
increased between 1999 and 2010 (Table 2) which is contrary to the expected behavior for a disk 
subjected to a brightening of the central AGN \cite[see for example, the discussion of 3C 332 in][]{Gez07}.
Second, there was likely no associated brightening of the central AGN as discussed further in the latter part of this section. Third, the disk is more appropriately described as a ring given the small ratio of outer to inner radius deduced from the model (Table 4). Fourth, the inner radius of the ring apparently decreased in size by ${\sim}$ 20\% between 1999 and 2010. Collectively, these observations suggest that we are witnessing a tidal disruption event unfolding in NGC 4203.
Tidally disrupted solar-type stars are unlikely candidates as the inner radius of the ring (1687 AU in 1999 and 1340 AU in 2010)\footnotemark \footnotetext{There are 0.609 AU/${r_g}$ in NGC 4203.} is about a factor of 10${^3}$ larger than the tidal radius expected for a 1 M${_\odot}$ main sequence star in the vicinity of a 6 ${\times}$ 10${^7}$ M${_\odot}$ BH. However, the tidal radius increases for objects of lower density. For example, the tidal radius, R${_t}$ at which co-rotating objects of density, ${\rho}$ are disrupted by a BH of mass, M${_\bullet}$ is given by

\begin{equation}
R_t =  0.78([\rho{_\odot}/\rho] [M{_\bullet}/10^7M{_\odot} ])^{1/3}           \rm ~~~ AU
\end{equation}

Thus, the average density of an object that will be tidally disrupted at a distance similar to the inner radius deduced for the ring
(Section 3), corresponds to ${\sim}$ 10${^{-9}}$ ${\rho_\odot}$, comparable to the average density expected for a late-type supergiant star, similar to Betelgeuse for example\footnotemark \footnotetext{The tidal radius increases for objects of lower density and the average density of Betelgeuse could range from 10${^{-8}}$ ${\rho_\odot}$ to 10${^{-13}}$ ${\rho_\odot}$ depending on its radius which could be as large as 134 AU \citep{Ker11}.}.  Consequently, it is quite plausible that the brightening of the double-peaked broad lines in NGC 4203 was caused by the ongoing tidal disruption of a red supergiant star which would have made ${\sim}$ 1.5 orbits in the time span of the {\it STIS} observations \cite[see][for evidence that red supergiant stars exist in S0 galaxies]{Dav10}.  

In the tidal disruption scenario, the average density of ionized gas responsible for the double-peaked
broad H${\alpha}$ emission line is likely to be comparable to the outer atmosphere of Betelgeuse for which n ${\sim}$ 5 ${\times}$ 10$^{11}$ cm$^{-3}$
\citep{Rav11}, high enough, perhaps, to explain the anomalous Balmer decrement noted in Section 2.0.4. The mass of gas can be estimated using 
equation 2, an effective recombination coefficient ${\alpha^{eff}_{H\alpha}}$ = 1.16 x 10$^{-13}$ cm${^{3}}$ s${^{-1}}$ and a luminosity $L (H{{\alpha}}$) = 1.38 x 10${^6}$ L${_{\sun}}$, based on the {\it difference} between the broad line flux measured in 2010 and 1999 which should more closely represent that produced by just the ring in 2010. The resulting mass\footnotemark \footnotetext{It is quite remarkable that the capabilities of {\it HST} are such that it can detect radiation from an ionized stellar contrail with a mass equivalent to 686 Moon masses in a galaxy 15 Mpc away.} of ionized gas corresponds to
${\sim}$ 2 ${\times}$ 10${^{-5}}$ M${_\odot}$ and is presumably strewn in a ring shaped contrail marking the trajectory of the supergiant much as envisaged by \citet[][see their Fig. 1]{Sco95} and \citet[][and references therein]{Bog04}. More recently \cite{Mai10} describe objects shaped like comets
orbiting and eclipsing the central X-ray source in NGC 1365. The defining characteristics of their ``comets'' -- a dense head
and low mass tail -- are intriguingly similar to the those proposed here for the supergiant producing the stellar-contrail in NGC 4203.

It was noted in section 4.3 that NGC 4203 suffers from an ionizing deficit that was particularly acute in 2010. The deficit could be alleviated if the AGN had brightened by a factor of 4 since 2003, but this is unlikely as the amplitude of the UV variability is ${\sim}$ 50\% at the shortest wavelength 
when it was measured by \cite{Mao05}. Furthermore, one can not appeal to a boost of X-rays as no X-ray variability has been detected \citep{Pia10, You11}. 
However, the broad H${\alpha}$ line emission observed in 2010 is dominated by the double-peaked broad line component that is attributed to the tidal disruption event. Thus, it is quite conceivable that ram pressure shock ionization produced by the interaction of the in-falling supergiant star with the ambient interstellar medium could provide an additional {\it mechanical} source of ionization that would help alleviate the ionizing deficit noted for NGC 4203. Note that ram pressure ionization is a very different mechanism 
for exciting the contrail than proposed by \citet{Rees88} which appeals to external {\it flares} generated by material as it is accreted onto the BH \citep[e.g.][]{Gez03}. Ram pressure ionization is produced {\it in-situ}. Although ram pressure ionization was discussed previously in the context of  
the tail of ionized gas leaving the Galactic Center M supergiant IRS 7 \citep{Ser91, Yus92} there is a need to explicitly model 
the tidal disruption of a supergiant star as existing studies have tended to focus on solar type stars \cite[e.g.][and references therein]{Bog04}. 

\section{Conclusions}

Spectroscopic observations with the {\it Hubble Space Telescope (HST)} have revealed a time variable double-peaked broad H${\alpha}$ emission line profile in NGC 4203
which has been successfully modeled as the combination of a {\it time variable disk plus a steady state inflow}. In this model the broad H${\alpha}$ line emission observed in 1999 is dominated by the inflow
which is large, ${\sim}$ 2 pc in diameter, and likely sustained by stellar mass loss. If the gas density is high, ${\geq}$ 10$^6$ cm$^{-3}$, as suggested by the absence of similarly broad 
[O I] and [O III] emission lines, then interpreting the broad H${\alpha}$ emission line observed in 1999 leads to a steady state inflow rate ${\sim}$ 2 ${\times}$ 10$^{-2}$ M${_{\sun}}$/yr which exceeds the requirement to explain the X-ray luminosity in terms of radiatively inefficient accretion by a factor of ${\sim}$ 6. The time variable double-peaked component of the broad H${\alpha}$ emission line, which is particularly impressive in the 2010 spectrum, has been modeled as an axis-symmetric disk, more appropriately described as a narrow ring, leading to the conclusion that it most likely represents the contrail of a supergiant star that is being tidally disrupted by the central BH. The AGN is apparently unable to sustain the ionization of the broad H${\alpha}$ emission line. The discrepancy is particularly acute in 2010 unless the AGN had brightened by a factor of 4 since 2003 which seems unlikely. Ram pressure shock ionization produced by the interaction of the in-falling supergiant star with the ambient interstellar medium could provide an additional mechanical source of ionization that would help alleviate the ionizing deficit noted for NGC 4203.




\acknowledgments
This research has made extensive use of the NASA Astrophysics Data System, the Atomic Line List, http://www.pa.uky.edu/~peter/newpage/
and a variety ${\bf STSDAS}$ tasks. Support for Program number HST-AR-11752.01-A was provided by NASA through a grant from the Space Telescope Science Institute, which is operated by the Association of Universities for
Research in Astronomy, Incorporated, under NASA contract NAS5-26555. The author thanks the anonymous referee for instructive comments that greatly improved the
presentation of this paper.



{\it Facilities:}  \facility{HST (STIS)}

\clearpage



\begin{figure}
\epsscale{1.0}
\begin{center}
\plotone{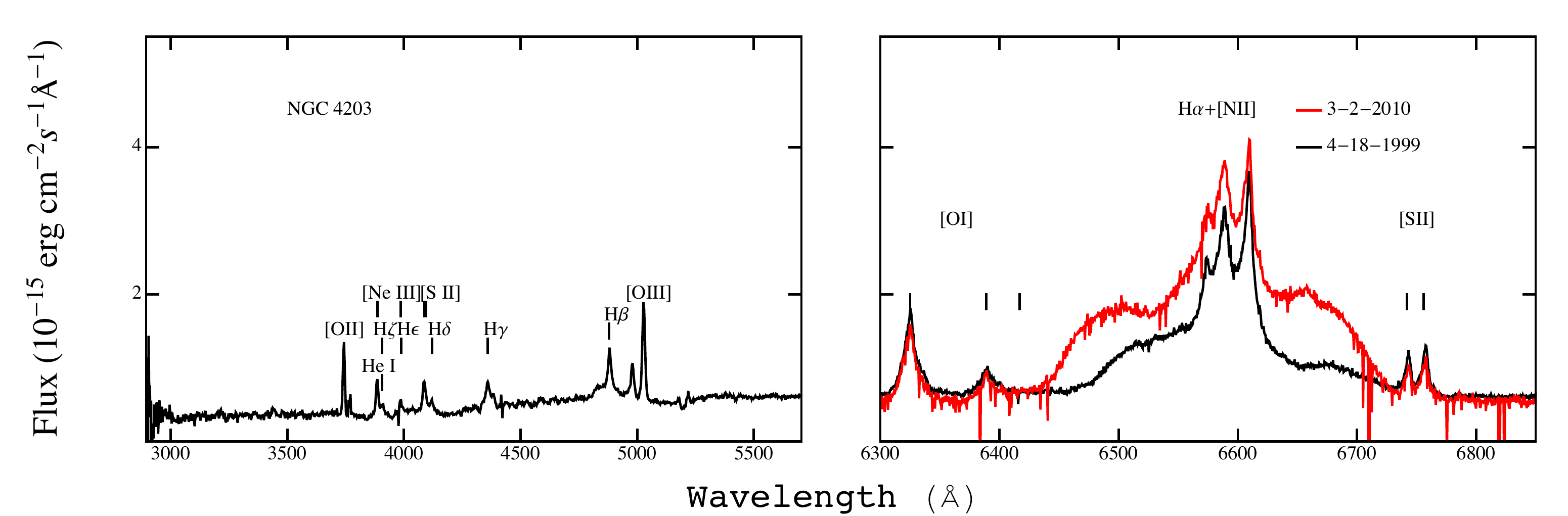}
\caption{{ Visual spectra  of NGC 4203 as seen through the following gratings:  {\sl Left panel}: G430L. {\sl Right panel}: G750M.
Red line shows data obtained under PID 11571. Black lines for both panels show data obtained under PID 7361.}}
\label{default}
\end{center}
\end{figure}

\clearpage

\begin{figure}
\epsscale{1.0}
\begin{center}
\plottwo{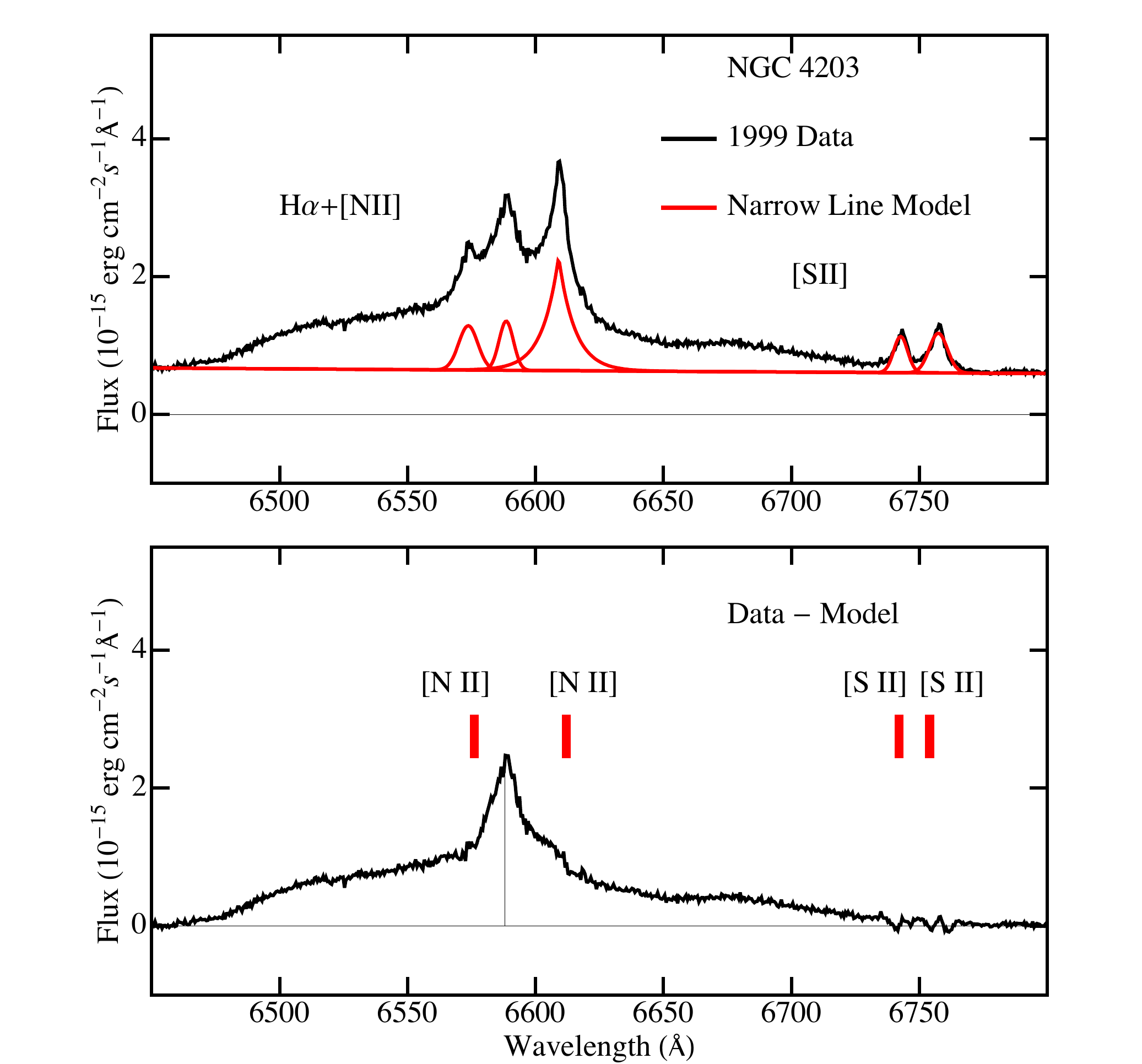}{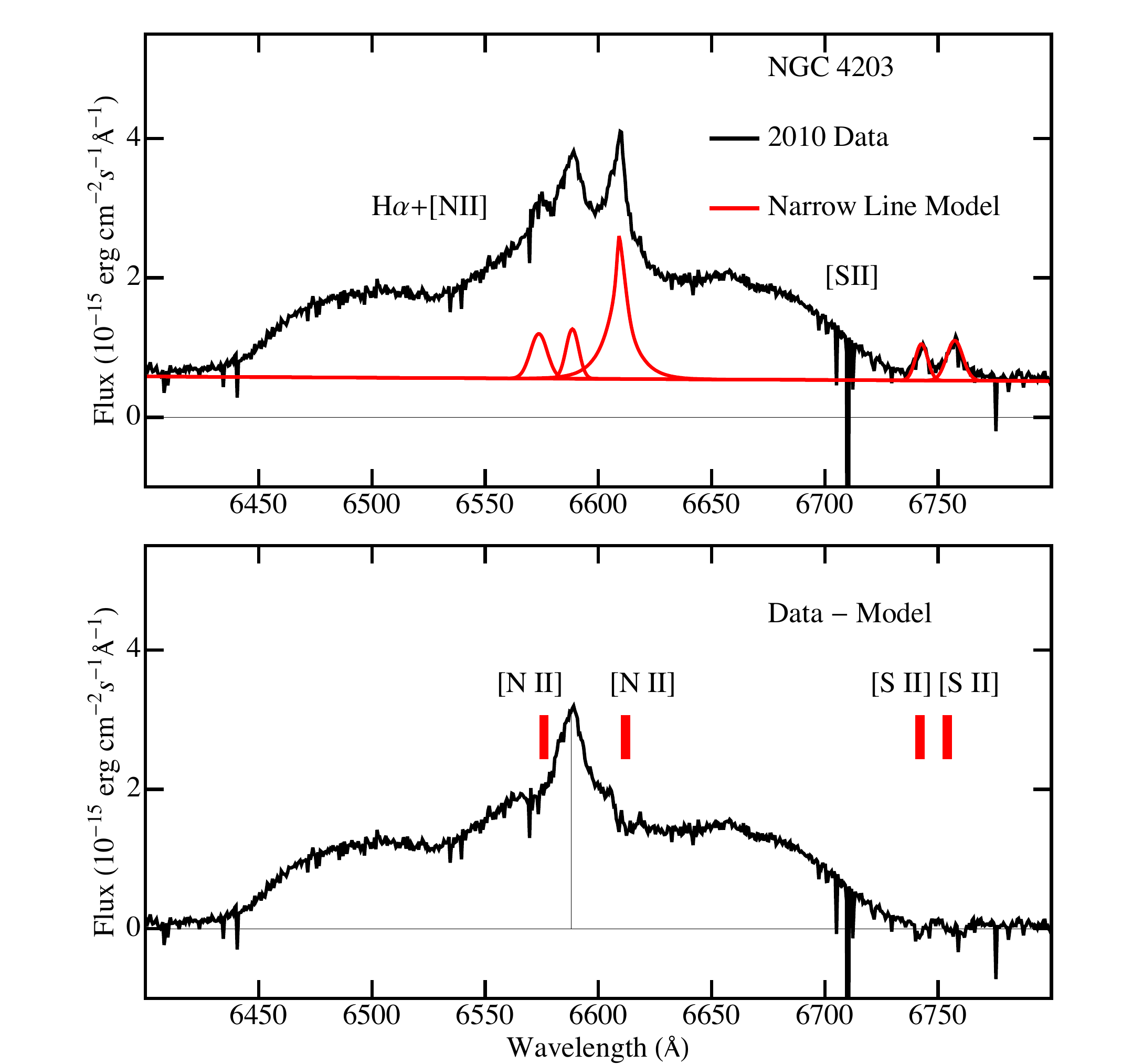}
\caption{{ Broad H${\alpha}$ emission line in NGC 4203 observed in 1999 ({\sl Left panels}) and 2010 ({\sl Right panels}). {\sl Top panels}: The observed spectrum is shown in black and a model for the forbidden lines is shown in red (see also Table 2). {\sl Lower panels}: The broad 
H${\alpha}$ emission line profile
after the forbidden lines have been subtracted. The central wavelengths of the subtracted lines are
indicated in red. The vertical black line corresponds to the observed (redshifted) central wavelength of the H${\alpha}$ line }}
\label{default}
\end{center}
\end{figure}

\clearpage

\begin{figure}
\epsscale{0.8}
\begin{center}
\plotone{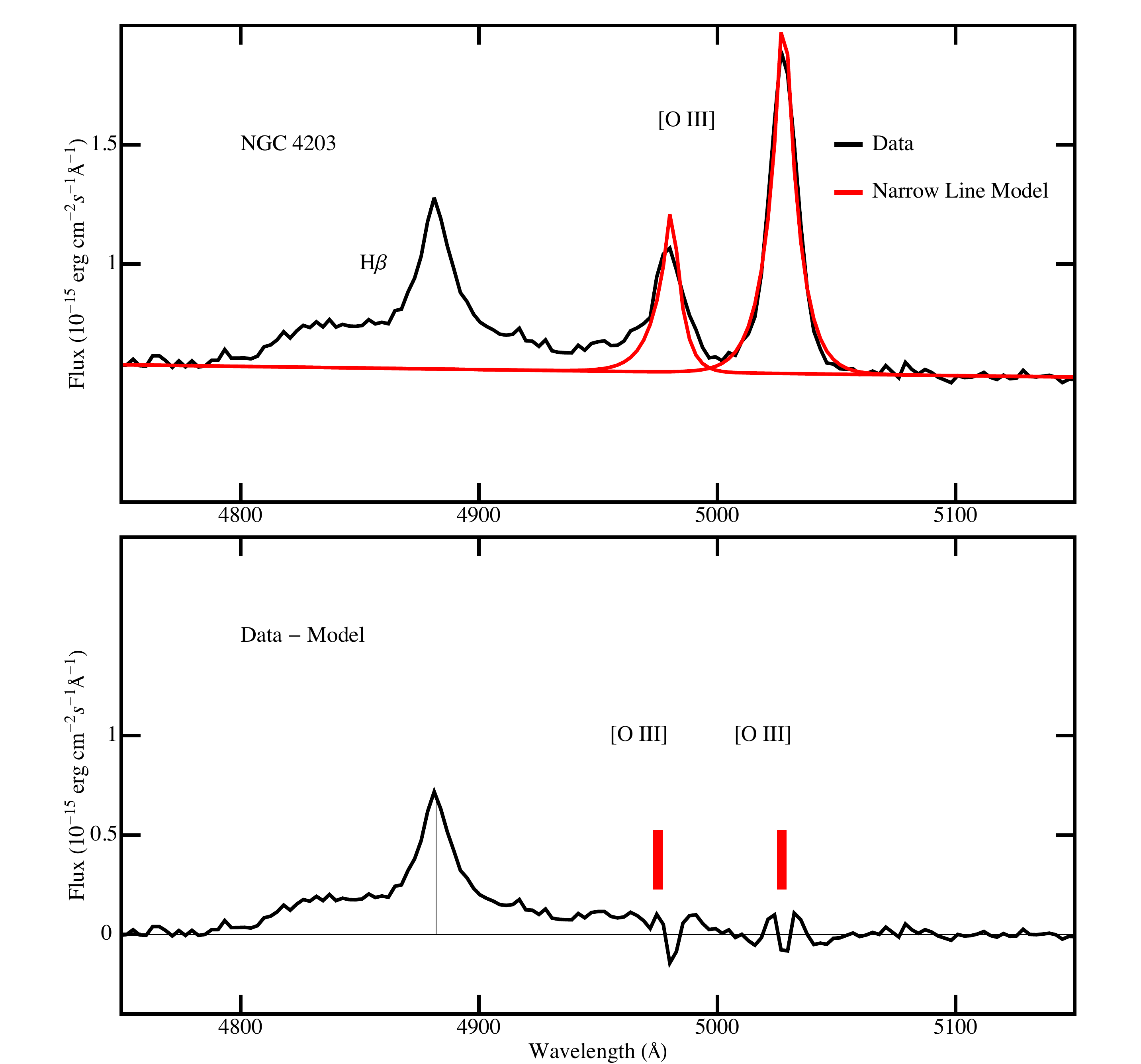}
\caption{{ Broad H${\beta}$ emission line in NGC 4203 observed in 1999. {\sl Top panel}: The observed spectrum
is shown in black and a model for the forbidden lines is shown in red (see also Table 2). {\sl Lower panel}: The broad 
H${\beta}$ emission line profile after the forbidden lines have been subtracted. The central wavelengths of the subtracted lines are
indicated in red. The vertical black line corresponds to the observed (redshifted) central wavelength of the H${\beta}$ line }}
\label{default}
\end{center}
\end{figure}

\clearpage

\begin{figure}
\epsscale{0.8}
\begin{center}
\plotone{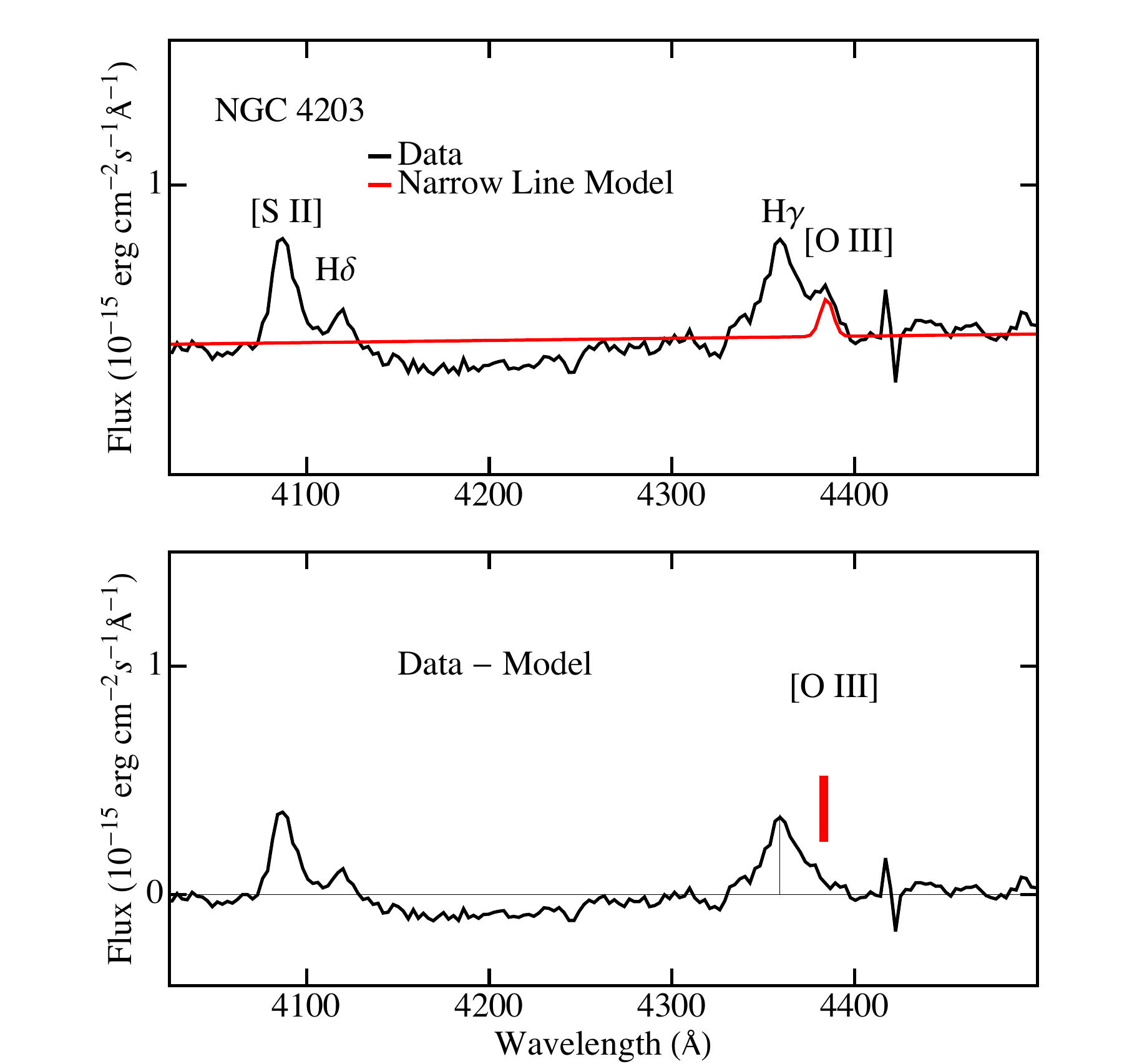}
\caption{{ Broad H${\gamma}$ emission line in NGC 4203 observed in 1999. {\sl Top panel}: The observed spectrum
is shown in black and a model for the forbidden line is shown in red (see also Table 2). {\sl Lower panel}: The broad 
H${\gamma}$ emission line profile
after the forbidden line has been subtracted. The central wavelength of the subtracted line is
indicated in red. The vertical black line corresponds to the observed (redshifted) central wavelength of the H${\gamma}$ line }}
\label{default}
\end{center}
\end{figure}

\clearpage

\begin{figure}
\epsscale{0.5}
\begin{center}
\plotone{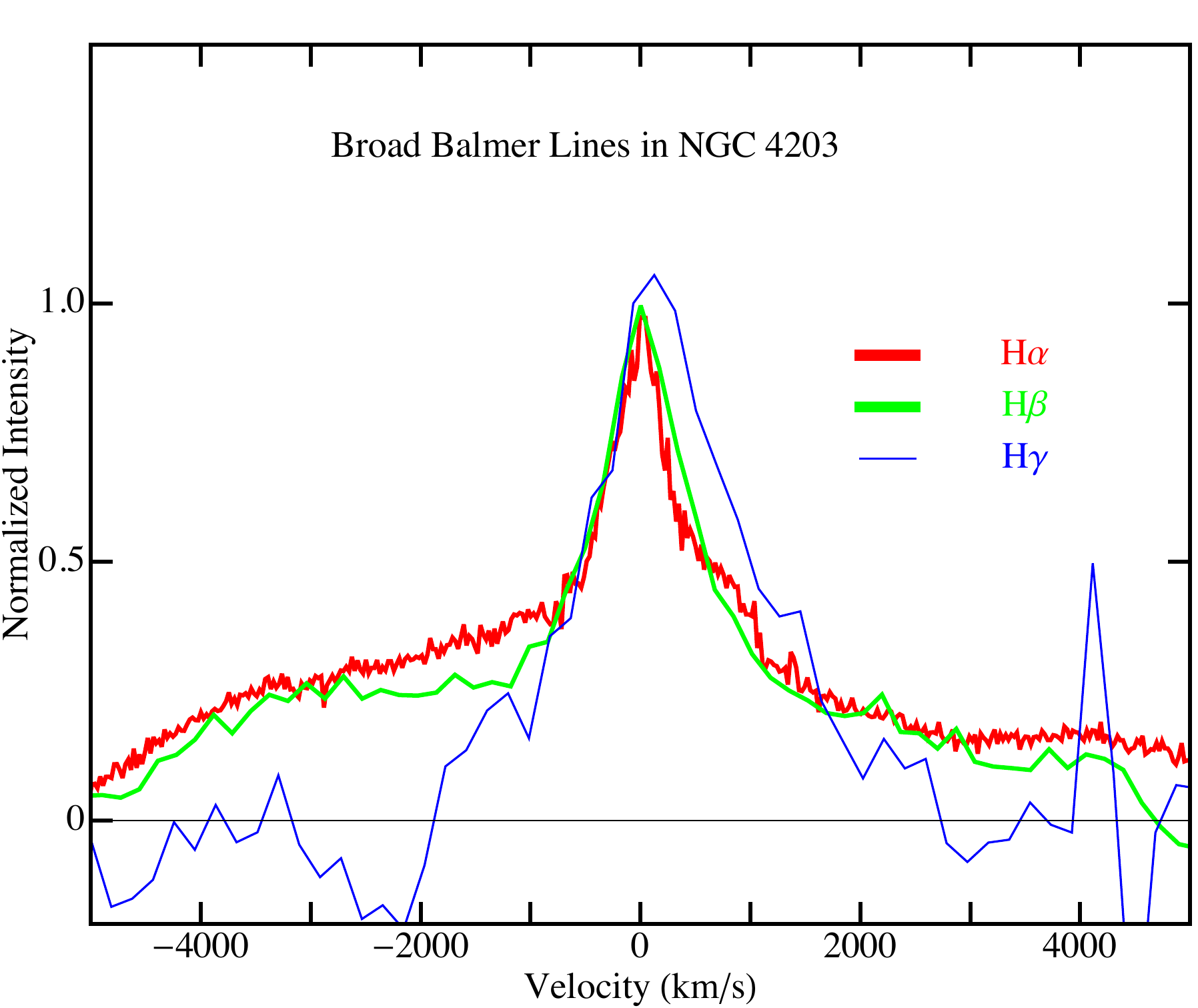}
\caption{{ (Left Panel). Comparison of the normalized broad
H${\alpha}$, H${\beta}$ and H${\gamma}$ emission lines in NGC 4203 observed in 1999.}}
\label{default}
\end{center}
\end{figure}

\clearpage

\begin{figure}
\epsscale{1.0}
\begin{center}
\plottwo{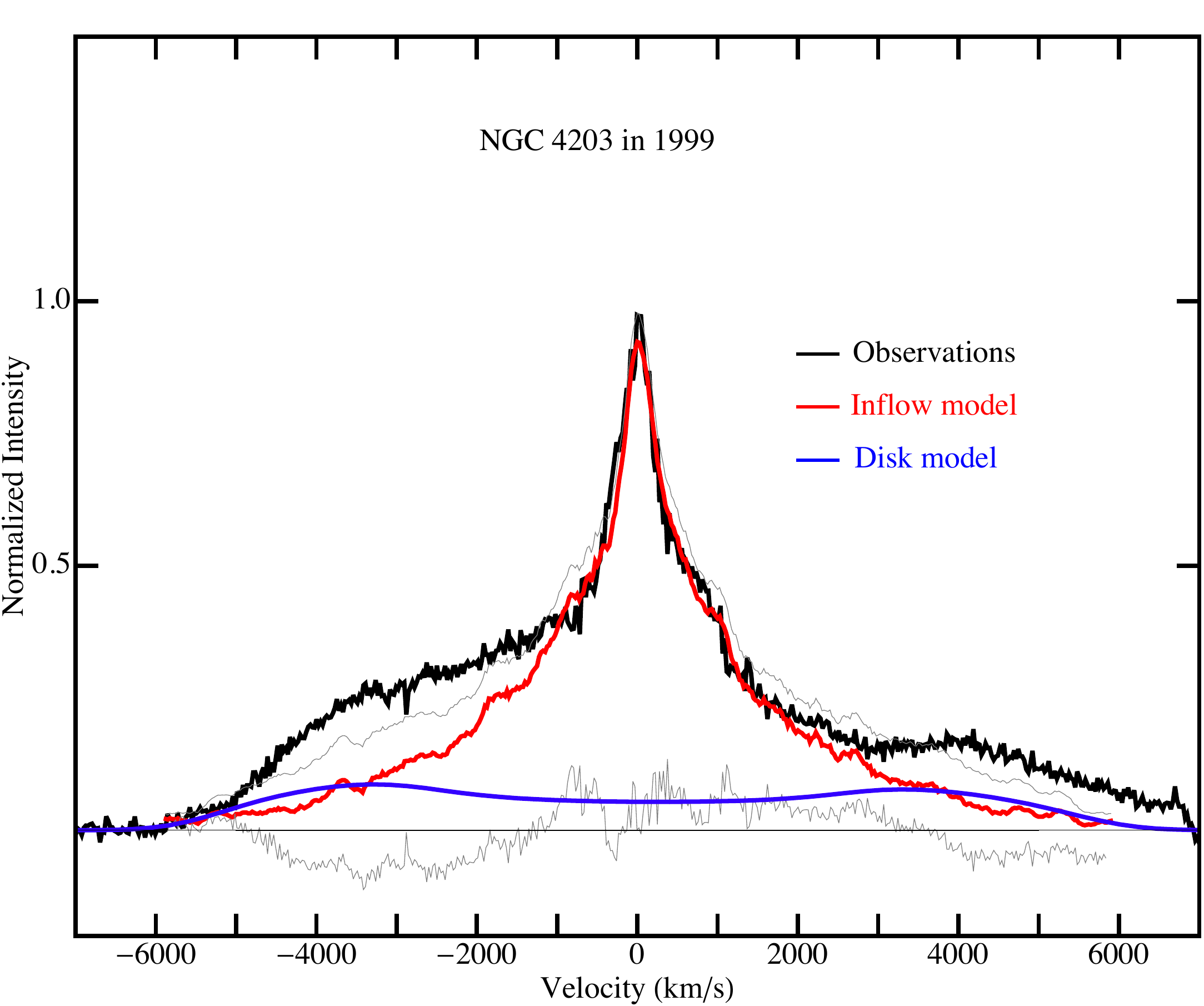}{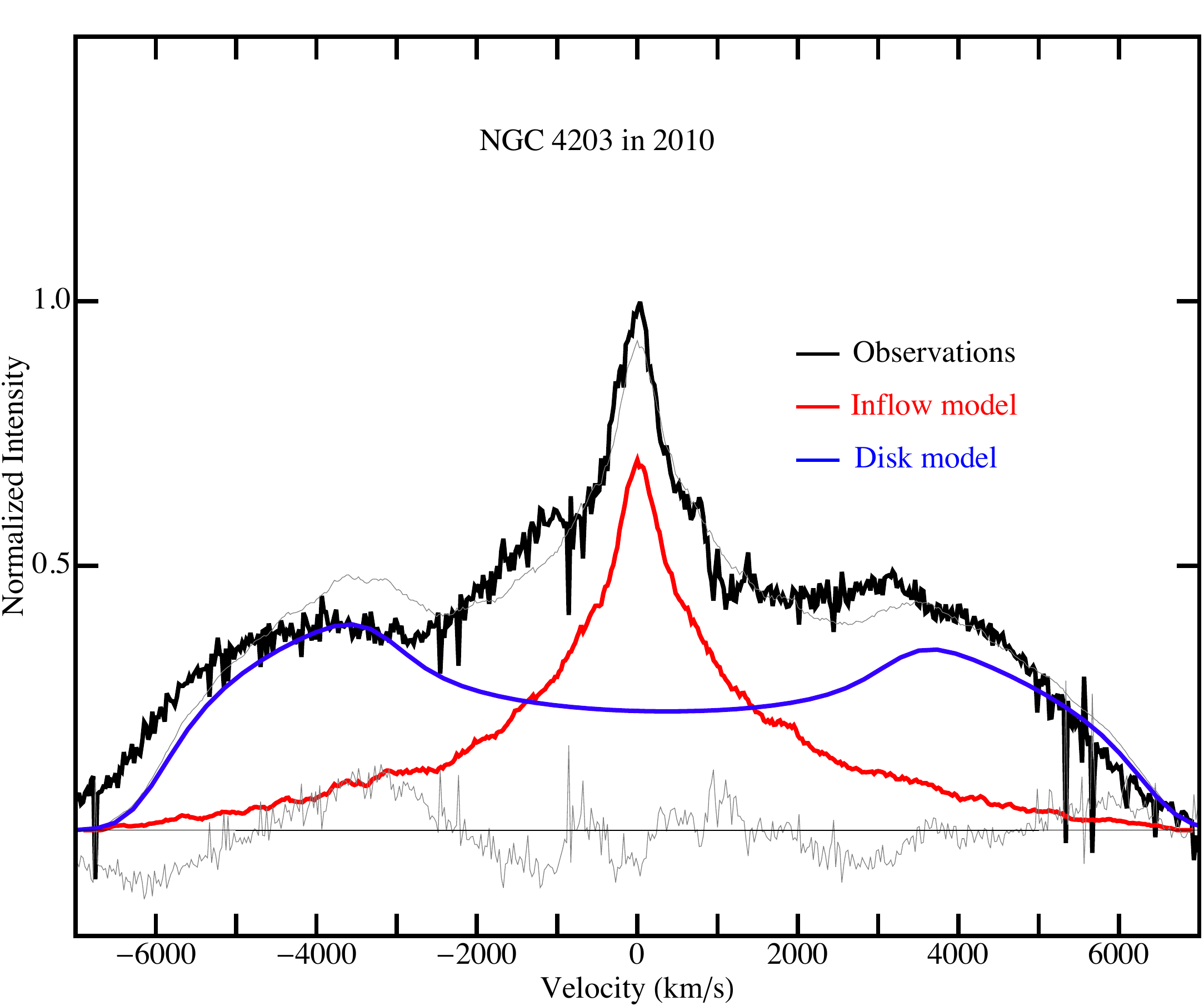}
\caption{{The broad H${\alpha}$ line observed in 1999 ({\sl Left panel}) and 2010 ({\sl Right panel}) modeled as a combination of a disk (blue line) and an inflow (red line). The thick black line represents the data. The sum of the model components and the residuals are plotted as thinner gray lines. These are just two of the many possible solutions
permitted by the ${\chi}^2$ surface illustrated in Fig. 7. The model disks plotted here both have an inclination of 82${^\circ}$ but different sizes and brightnesses (see Table 6) and represent the solutions at the tip and the tail of the arrow plotted in Fig. 7.}}
\label{default}
\end{center}
\end{figure}

\clearpage

\begin{figure}
\epsscale{1.0}
\begin{center}
\plotone{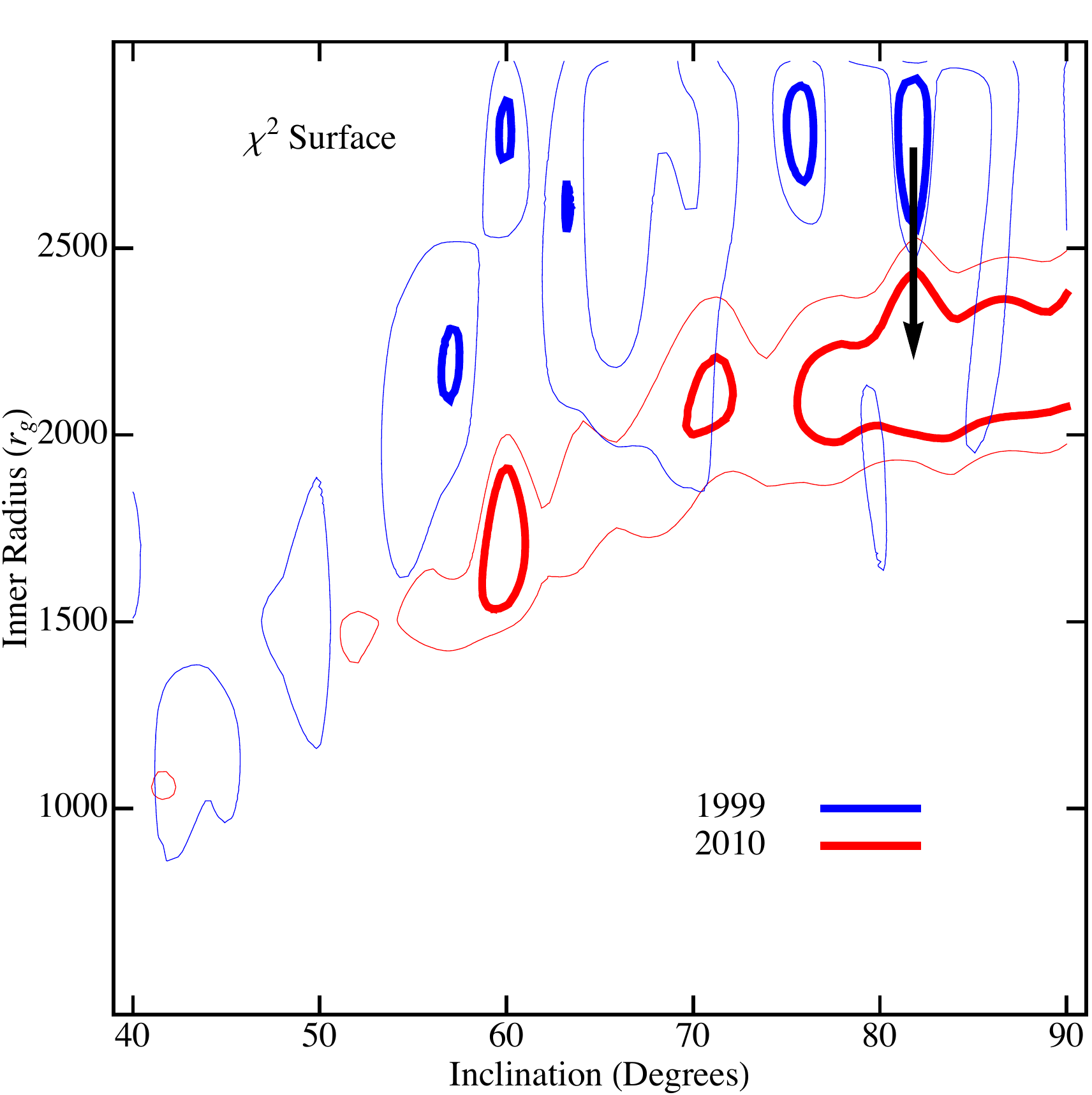}
\caption{{Contour plot of the reduced ${\chi}^2$ surface constraining the model disk parameters that best represent the NGC 4203 spectra observed in 2010 (red contours)
and 1999 (blue contours). The thicker lines define the ${\chi}^2_{min}$ contour. The thinner lines define the ${\chi}^2_{min}$ + 2${\sigma}$ contour. There are a number of degenerate solutions for each set of observations. However, the red and blue ${\chi}^2_{min}$ contours do not overlap suggesting that there
is no single disk model that can simultaneously explain both the 1999 and 2010 observations. The arrow represents the ${\it least}$ size evolution solution (see Section 4.7 for details). }}
\label{default}
\end{center}
\end{figure}

\clearpage

\begin{deluxetable}{ccccccccc}
\tabletypesize{\scriptsize}
\tablecaption{NGC 4203 Spectral Datasets\label{tbl-2}}
\tablewidth{0pt}
\tablehead{
\colhead{PID} & \colhead{Observation Date} & \colhead{Grating} & \colhead{Spectral Range} & \colhead{Slit} & \colhead{Dispersion} & \colhead{Plate Scale} & \colhead{Integration Time} & \colhead{Datasets}   \\
\colhead{} & \colhead{} &  \colhead{} & \colhead{\AA}  & \colhead{arc sec} & \colhead{\AA/pixel}& \colhead{arc sec/pixel} & \colhead{s} & \colhead{}\\
\colhead{(1)} & \colhead{(2)} &  \colhead{(3)} & \colhead{(4)}  & \colhead{(5)} & \colhead{(6)} & \colhead{(7)} & \colhead{(8)} & \colhead{(9)} \\
}
\startdata
7361 & 4-18-1999 &  G750M  & 6295 - 6867 & 52 x 0.2 & 0.56 & 0.05 & 900 & o4e010010 \\
7361 & 4-18-1999 &  G750M  & 6295 - 6867 & 52 x 0.2 & 0.56 & 0.05 & 979 & o4e010020 \\
7361 & 4-18-1999 &  G750M  & 6295 - 6867 & 52 x 0.2 & 0.56 & 0.05 & 900 & o4e010030 \\
7361 & 4-18-1999 &  G430L  & 2900 - 5700 & 52 x 0.2 & 2.73 & 0.05 & 730 & o4e010040 \\
7361 & 4-18-1999 &  G430L  & 2900 - 5700 & 52 x 0.2 & 2.73 & 0.05 & 900 & o4e010050 \\
11571 & 3-2-2010 &  G750M  & 6295 - 6867 & 52 x 0.1 & 0.56 & 0.05 & 860 & ob3i01010 \\
11571 & 3-2-2010 &  G750M  & 6295 - 6867 & 52 x 0.1 & 0.56 & 0.05 & 860 & ob3i01020 \\
11571 & 3-2-2010 &  G750M  & 6295 - 6867 & 52 x 0.1 & 0.56 & 0.05 & 728 & ob3i01030 \\
\enddata

\end{deluxetable}

\clearpage

\begin{deluxetable}{cccc}
\tabletypesize{\scriptsize}
\tablecaption{Emission Line Parameters for the G750M Nuclear Spectrum Obtained 4-18-1999\tablenotemark{a}}
\tablewidth{0pt}
\tablehead{
\colhead{Line} & \colhead{Central Wavelength\tablenotemark{b}} & \colhead{Flux} & \colhead{FWHM}   \\
\colhead{} & \colhead{\AA} &  \colhead{10$^{-15}$ erg cm$^{-2}$ s$^{-1}$} & \colhead{kms$^{-1}$} \\
\colhead{(1)} & \colhead{(2)} &  \colhead{(3)} & \colhead{(4)} \\
}
\startdata
$\textrm{[O I]}$ &  6325  ${\pm}$ 1 & 12.3 ${\pm}$ 2.0 &  630 ${\pm}$  131 \\
$\textrm{[O I]}$  &  6390  ${\pm}$ 3 & 3.9 ${\pm}$ 2.1 &  649 ${\pm}$  440 \\
$\textrm{[N II]}$ &  6574 ${\pm}$ 3   & 6.0 ${\pm}$  3.2  &  400 ${\pm}$ 147  \\
H${\alpha}$ (broad)\tablenotemark{c} & 6588  & 168 ${\pm}$ 2 & 1300 ${\pm}$ 100 \\
H${\alpha}$ (broad)\tablenotemark{d} & 6588  & 372 ${\pm}$ 4 & 2667 ${\pm}$ 100 \\
H${\alpha}$ (narrow) & 6588 ${\pm}$ 3  & 5.0 ${\pm}$ 2.0 & 300 ${\pm}$ 85  \\
$\textrm{[N II]}$ & 6609 ${\pm}$ 1  & 21.2 ${\pm}$ 2.5\tablenotemark{e} & 400 ${\pm}$ 28  \\
$\textrm{[S II]}$ & 6742 ${\pm}$ 1  & 3.5 ${\pm}$ 0.7 & 283 ${\pm}$ 75 \\
$\textrm{[S II]}$ & 6757 ${\pm}$ 1 & 5.1  ${\pm}$ 0.8 & 371 ${\pm}$ 75 \\
\enddata
\tablenotetext{a}{Measured within a 0.2{\arcsec}  x 0.35{\arcsec}  aperture. Continuum subtracted but not corrected for dust extinction. }
\tablenotetext{b}{Observed wavelength}
\tablenotetext{c}{1999}
\tablenotetext{d}{2010, measured in a 0.1{\arcsec}  x 0.35{\arcsec}  aperture}
\tablenotetext{e} {The [N II] emission line flux is chosen so as to not over-subtract the broad H${\alpha}$ emission line profile. The
[N II] line shape is modeled using two components; a logarithmic profile of width 400 km/s 
and a gaussian of width 130 km/s each contributing 90\% and 10\% of the emission line flux, respectively. }
\end{deluxetable}

\clearpage

\begin{deluxetable}{cccc}
\tabletypesize{\scriptsize}
\tablecaption{Emission Line Parameters for the G430L Nuclear Spectrum Obtained 4-18-1999\tablenotemark{a}}
\tablewidth{0pt}
\tablehead{
\colhead{Line} & \colhead{Central Wavelength\tablenotemark{b}} & \colhead{Flux\tablenotemark{c}} & \colhead{FWHM}   \\
\colhead{} & \colhead{\AA} &  \colhead{10$^{-14}$ erg cm$^{-2}$ s$^{-1}$} & \colhead{kms$^{-1}$} \\
\colhead{(1)} & \colhead{(2)} &  \colhead{(3)} & \colhead{(4)} \\
}
\startdata
$\textrm{[O II]}$& 3742  ${\pm}$ 1 & 0.96 ${\pm}$ 0.02 &  764 ${\pm}$  24 \\
H${\gamma}$ (broad) & 4360 &  0.8 ${\pm}$ 0.1 & 1770 ${\pm}$ 200 \\
$\textrm{[O III]}$\tablenotemark{d} &  4384  ${\pm}$ 1 &  ${\leq}$ 0.1  &  570 \\
H${\beta}$ (broad) & 4882  &  3.33 ${\pm}$ 0.04 & 1263 ${\pm}$ 100 \\
$\textrm{[O III]}$\tablenotemark{e}&  4981  ${\pm}$ 1 & 0.9 ${\pm}$ 0.1 &  570 \\
$\textrm{[O III]}$\tablenotemark{e}&  5028  ${\pm}$ 1 & 2.2 ${\pm}$ 0.2 &  570 ${\pm}$  64 \\

\enddata
\tablenotetext{a}{Table entries that do not include uncertainties are fixed parameters.}
\tablenotetext{b}{Observed wavelength}
\tablenotetext{c}{Measured within
a 0.2{\arcsec}  x 0.35{\arcsec}  aperture. Continuum subtracted but not corrected for dust extinction. }
\tablenotetext{d} {The [O III] emission line flux is chosen so as to not over-subtract the broad H${\gamma}$ emission line profile}
\tablenotetext{e} {Adopting a logarithmic profile shape.}

\end{deluxetable}

\clearpage

\begin{deluxetable}{lc}
\tabletypesize{\scriptsize}
\tablewidth{0in}
\tablecaption{Inflow Plus Disk Model Fixed Parameters} 
\tablehead{
\colhead{Parameter} &  \colhead{Value} \\

}
\startdata
Disk outer to inner radius ratio, $\xi_o$/$\xi_i$               &    3.75  \\
Disk Broadening parameter, ${\sigma}$       & 350 km s$^{-1}$ \\
Disk Emissivity index, q & 3.0 \\
Inflow inner radius,  $r_{inner}$ & 0.01 pc\\
Inflow outer radius, $r_{outer}$ & 1 pc \\
Black hole mass, M${_{\bullet}}$ & 6 ${\times}$ 10${^7}$ M${_{\odot}}$ \\
\enddata

\end{deluxetable}

\clearpage

\begin{deluxetable}{lcc}
\tabletypesize{\scriptsize}
\tablewidth{0in}
\tablecaption{ Parameters for Computing ${\chi}_{red}^2$ }
\tablehead{
\colhead{} &  \colhead{Observation} & \colhead{Date}  \\
\colhead{Parameter} & \colhead{1999} & \colhead{2010} \\

}
\startdata
Summation performed over ${j}$ points  & 461       & 552  \\
Velocity range of summation, km/s  & -5860 to 5860 & -7010 to 7010 \\
\tablenotemark{a}Degrees of freedom, ${\nu}$  &  452    & 543 \\
Minimum ${\chi}_{red}^2$  & 6 & 1 \\
Inflow/Disk Intensity ratio normalized at v=0  & 10 & 2.75 \\

\enddata

\tablenotetext{a}{${\nu}$ = ${j}$ data points - 8 parameters - 1}

\end{deluxetable}

\clearpage


\begin{thebibliography}{}

\bibitem[Alexander \& Netzer (1994)]{Ale94} Alexander, T., \& Netzer, H., 1994, \mnras, 270, 781

\bibitem[Barth et al. (2001)]{Bar01} Barth, Aaron J., Ho, L. C., Filippenko, A. V., Rix, H-W., Sargent, W. L. W., 2001, \apj, 546, 205

\bibitem[Barth, Ho, \& Sargent (2002)]{Bar02} Barth, A. J., Ho, L. C., \&  Sargent, W. L. W., 2002, \aj, 124, 2607

\bibitem[Bassin \& Bonatto (2003)]{Bas03} Bassin, M.,  \& Bonatto, Ch., 2003, \aap, 410, 803

\bibitem[Beifiori et al. (2009)]{Bei09} Beifiori, A., et al., 2009, \apj, 692, 856

\bibitem[Bogdanovi\'c et al. (2004)]{Bog04} Bogdanovi\'c, T., Eracleous, M., Mahadevan, S., Sigurdsson, S., Laguna, P., 2004, \apj, 610, 707

\bibitem[Bondi (1952)]{Bon52} Bondi, H., 1952, \mnras, 112, 195

\bibitem[Bower et al. (1996)]{Bow96} Bower, G.A., Wilson, A.S., Heckman, T.M., \& Richstone, D.O.,1996, \aj, 111, 1901

\bibitem[Chen \& Halpern (1989)]{Che89} Chen, K., \& Halpern, J.P., 1989, \apj, 344, 115  

\bibitem[Davidge (2010)]{Dav10} Davidge, T.J., 2010, \aj, 139, 680

\bibitem[Devereux (2011)] {Dev11} Devereux, N., 2011, \apj, 727, 93

\bibitem[Devereux, Ford, Tsvetanov, \& Jacoby (2003)] {Dev03} Devereux, N. A., Ford, H. C., Tsvetanov, Z., \&  Jacoby, G., 2003 \aj, 125, 1226

\bibitem[Devereux \& Shearer (2007)]{Dev07} Devereux, N.,  \&   Shearer, A., 2007, \apj, 671, 118

\bibitem[Down et al. (2010)]{Dow10} Down, E. J., Rawlings, S., Sivia, D. S., \& Baker, J. C., 2010, \mnras, 402, 633

\bibitem[Eracleous \& Halpern (1994)]{Era94} Eracleous, M., \& Halpern J.P., 1994, \apjs, 90, 1

\bibitem[Eracleous \&  Halpern (2001)]{Era01} Eracleous, M., \& Halpern, J. P., 2001, \apj, 554, 240

\bibitem[Filippenko \& Hapern (1984)]{Fil84} Filippenko, A.V., \& Halpern, J.P., 1984, \apjs, 285, 458

\bibitem[Gezari et al. (2003)]{Gez03} Gezari, S., Komossa, S., Grupe, D., \& Leighly, K. M., 2003, \apj, 592, 42

\bibitem[Gezari, Halpern \& Eracleous (2007)]{Gez07} Gezari, S., Halpern, J.P., \& Eracleous, M., 2007, \apjs, 169, 167 

\bibitem[ Gonzales Delgado et al.  (2004)]{GD04} Delgado Gonzales et al., 2004, \apj, 605, 127

\bibitem[Greene \& Ho (2005)]{Gre05} Greene, J.E., \& Ho, L.C., 2005, \apj, 630, 122

\bibitem[Heckman (1980)]{Hec80} Heckman, T. M., (1980) \aap, 87, 152

\bibitem[Ho (2009)]{Ho09} Ho, L.C., 2009, \apj, 699, 626

\bibitem[Ho, Filippenko, \& Sargent  (1997)]{Ho97a} Ho, L.C., Filippenko, A.V.,  \& Sargent, W.L.W.,  1997, \apjs, 112, 315

\bibitem[Ho, Filippenko \& Sargent (1996)]{Ho96} Ho, L.C., Filippenko, A.V., \& Sargent, W.L.W., 1996, \apj, 462, 183

\bibitem[Ho, Filippenko, Sargent  \& Peng (1997)]{Ho97b} Ho, L.C., Filippenko, A.V.,  Sargent, W.L.W., \& Peng, C.Y., 1997, \apjs, 112, 391

\bibitem[Ho et al. (2000)]{Ho00} Ho, L. C., Rudnick, G., Rix, H-W., Shields, J. C., McIntosh, D. H.,  Filippenko, A. V., Sargent, W. L. W., Eracleous, M., 2000, \apj, 541, 120

\bibitem[Kaspi et al. (2005)]{Kas05} Kaspi, S., Maoz, D., Netzer, H.,Peterson, B.M., Vestergaard, M., \& Jannuzi, B.T., 2005, \apj, 629, 61

\bibitem[Kervella et al. (2005)]{Ker11}  Kervella, P., et al., 2011, \aap, 531, 117

\bibitem[Kewley et al. (2006)]{Kew06} Kewley, L.J., Groves, B., Kauffmann, G., \& Heckman, T., 2006, \mnras, 372, 961

\bibitem[Lacy, Achtermann \& Serabyn, (1991)]{Lac91} Lacy, J. H., Achtermann, J. M., \& Serabyn, E., 1991, \apj, 380, 71
	
\bibitem[Lewis, Eracleous, \& Storchi-Bergmann (2010)]{Lew10} Lewis, K.T.,  Eracleous, M.,  \& Storchi-Bergmann, T., 2010, \apjs, 187, 416

\bibitem[Lewis \& Eracleous (2006)]{Lew06} Lewis, K.T.,  \& Eracleous, M., 2006, \apj, 642, 711

\bibitem[Maiolino et al. (2010)]{Mai10} Maiolino, R., et al., 2010, \aap, 517, 47

\bibitem[Maoz et al. (1998)]{Mao98} Maoz, D., Koratkar, A., Shields, J.C., Ho, L.C., Filippenko, A.V., \& Sternberg, A., 1998, \aj, 116, 55

\bibitem[Maoz et al. (2005)]{Mao05} Maoz, D., Nagar, N.M., Falcke, H., \& Wilson, A., 2004, \apj, 625, 699

\bibitem[Maoz (2007)]{Mao07} Maoz, D., 2007, \mnras, 377, 1696

\bibitem[Merloni, Heinz, \& Di Matteo (2003)]{Mer03}  Merloni, A., Heinz, S., \& Di Matteo, T., 2003, \mnras, 345, 1057

\bibitem[Nagar, Falcke, Wilson \& Ulvestad, (2002)]{Nag02} Nagar, N. M., Falcke, H., Wilson, A. S., \&  Ulvestad, J. S., 2002, \aap, 392, 53

\bibitem[Noel--Storr et al. (2003)]{NS03} Noel--Storr et al., 2003, \apjs, 148, 419

\bibitem[Peterson (1993)]{Pet93} Peterson, B., 1993, \pasp, 105, 247

\bibitem[Peterson (2001)]{Pet01} Peterson, B., 2001, in Advanced Lectures on the Starburst-AGN Connection,    ed. I. Aretxaga, D. Kunth, \&  R. Mujica (Singapore: World Sci.), 3  

\bibitem[Pian et al.  (2010)]{Pia10} Pian, E.,  Romano, P.,   Maoz, D., Cucchiara, A., Pagani, C., \& La Parola, V., 2010 \mnras, 401, 677

\bibitem[Ravi et al. (2011)]{Rav11} Ravi, V., et al., 2011, arXiv:1105.3273

\bibitem[Rees (1988)]{Rees88} Rees, M. J., \nat, 523, 333

\bibitem[Proffitt et al. (2010)]{Pro10} Proffitt, C., et al., 2010, STIS Instrument Handbook, Version 9.0, (Baltimore: STScI).

\bibitem[Sarzi et al. (2001)]{Sar01} Sarzi, M., et al., 2001, \apj, 550, 65

\bibitem[Sarzi et al. (2002)]{Sar02} Sarzi, M., et al., 2002, \apj, 567, 237

\bibitem[Sarzi et al. (2005)]{Sar05} Sarzi, M., et al., 2005, \apj, 628, 169

\bibitem[Serabyn, Lacy, \& Achtermann (1991)]{Ser91} Serabyn, E., Lacy, J. H., \& Achtermann, J. M., 1991, \apj, 378, 557

\bibitem[Seyfert (1943)]{Sey43} Seyfert, C.K., 1943, \apj, 97, 28 

\bibitem[Scoville \& Norman (1995)]{Sco95} Scoville, N., \& Norman, C., 1995, \apj, 451, 510

\bibitem[Shields et~al.~(2000)]{Shi00} {Shields}, J. C., et~al. 2000, \apjl, 534, 27

\bibitem[Shields et~al.~(2007)]{Shi07} {Shields}, J. C., et~al. 2000, \apjl, 654, 125

\bibitem[Storchi-Bergmann (2009)]{Sto09} {Storchi-Bergmann}, The Monster's Fiery Breath: Feedback in Galaxies, Groups, and Clusters. AIP Conference Proceedings, Volume 1201, pp. 88-91 (2009).

\bibitem[Storchi-Bergmann, Baldwin \& Wilson (1993)]{Sto93} Storchi-Bergmann, T., Baldwin, J. A., Wilson, A. S., 1993, \apjl, 410, 11 

\bibitem[Storchi-Bergmann et al., (1995)]{Sto95} Storchi-Bergmann, T., Eracleous, M.,  Livio, M., Wilson, A. S., Filippenko, A. V., Halpern, J. P., 1995, \apj, 443, 617

\bibitem[Storchi-Bergmann et al., (1997)]{Sto97} Storchi-Bergmann, T., Eracleous, M.,  Ruiz, M.T., Livio, M., Wilson, A. S., Filippenko, A. V., 1997, \apj, 489, 87

\bibitem[Tonry et~al. (2001)]{Ton01} Tonry, J., et~al., 2001, \apj, 546, 681

\bibitem[Ulvestad \& Ho (2001)]{Ulv01} Ulvestad, J.S.,  \& Ho, L.C., 2001, \apjl, 562, 133

\bibitem[Younes, Porquet, Sabra, \& Reeves (2011)]{You11} Younes, G., Porquet, D., Sabra, B., \&  Reeves, J. N., 2011, \aap, 530, 149

\bibitem[Yusef-Zadeh \& Melia (1992)]{Yus92} Yusef-Zadeh, F., Melia, F., 1992, \apjl, 385,41


\end{thebibliography}
\end{document}